\documentclass{ijmpb}

\usepackage{amssymb}
\usepackage{slashed}
\usepackage{graphics}
\usepackage{epsfig}
\usepackage{bm}
\usepackage{color}
\usepackage[colorlinks,citecolor=blue,linkcolor=blue]{hyperref}
\usepackage{soul}

\newcommand{\be}{\begin{equation}}
\newcommand{\ee}{\end{equation}}
\newcommand{\bea}{\begin{eqnarray}}
\newcommand{\eea}{\end{eqnarray}}

\begin{document}

\markboth{Mizher, Raya, Villavicencio}
{Electric current generation in distorted graphene}

%
\catchline{}{}{}{}{}
%

\title{ELECTRIC CURRENT GENERATION IN DISTORTED GRAPHENE}

\author{ANA JULIA MIZHER}

\address{Instituto de Ciencias Nucleares, Universidad Nacional Aut\'onoma de M\'exico\\
Apartado Postal 70-543,   M\'exico Distrito Federal 04510, Mexico}

\author{ALFREDO RAYA}
\address{Instituto de F\'isica y Matem\'aticas, Universidad Michoacana de San Nicol\'as de Hidalgo\\
Edificio C-3, Ciudad Universitaria, C.P. 58040,
Morelia, Michoac\'an, Mexico}

\author{CRISTI\'AN VILLAVICENCIO}
\address{Departamento de Ciencias B\'asicas, Universidad del B\'io-B\'io\\
Casilla 447,
Chill\'an, Chile}

\maketitle

\begin{history}
\end{history}

\begin{abstract}
Graphene-like materials can be effectively described by  Quantum Electrodynamics in (2+1)-dimensions. 
In a pristine state, these systems exhibit a symmetry between the nonequivalent Dirac points in the honeycomb lattice. 
Realistic samples which include distortions and crystaline anisotropies are considered through mass gaps of topological and dynamical nature.
In this work we show that the incorporation of an in-plane uniform external magnetic field on this pseudochiral asymmetric configuration  generates a non-dissipative electric current aligned with the magnetic field: The pseudo chiral magnetic effect. This scenario resembles the chiral magnetic effect in Quantum Chromodynamics.
\end{abstract}

\keywords{Graphene; Magnetic field; Chiral magnetic effect}

\section{Introduction}

In less than a decade from the emergence of the so-called {\em Dirac and Weyl 
Materials} 
--among which graphene\cite{graphene,Gusynin:2007ix} and more recently, topological insulators\cite{topins}, have attracted considerable attention, 
a renewed interest within both the elementary particle and condensed matter physics has developed around the behavior of planar  fermions, which have transited from being toy models of Quantum Chromodynamics (QCD) under extreme conditions to actual players of a revolutionary era for fundamental physics and technological application prospects.

Graphene and related materials are composed by a single layer of atoms tightly packed into a two-dimensional honeycomb array and therefore can be efficiently described by tight-binding models. 
In the continuum limit, these models can be mapped into the Hamiltonian of $(2+1)$-dimensional quantum electrodynamics (QED$_3$) with massless Dirac fermions.\cite{Gusynin:2007ix}
A variety of {\em traditional} condensed matter phenomena found an effective description in terms of the QED$_3$ degrees of freedom, including high-$T_c$ superconductivity\cite{supercon} and quantum Hall effect.\cite{QHE} 
It was, however, the gapless nature of the charge carriers in 
graphene at low energy, which around the Dirac points of the Brillouin zone of 
the honeycomb lattice exhibit a linear dispersion relation what boosted the 
interest on the properties of ``relativistic'' planar  fermions in a condensed matter 
environment.

Although being Abelian, QED$_3$  exhibits similar features to non-Abelian gauge theories, hence establishing a link between particle physics and condensed matter systems. 
Therefore, QED$_3$ opens the possibility to explore phenomena which are either inaccessible, due to energy limitations, or hard to measure in a particle physics experiment. 
Indeed, it is known that at very high temperatures, a non-Abelian gauge theory coupled to $N_f$ fermion families in (3+1)-space--time dimensions experiences a dimensional reduction to an effective (2+1)-dimensional theory, which further ``abelianizes'' if $N_f$ is large enough\cite{pisarski}; non-abelian interactions are suppressed by a factor of $N_f^{-1}$. 
This fact makes QED$_3$ an effective version of QCD, which also exhibits 
important non-perturbative phenomena like confinement and dynamical chiral 
symmetry breaking.\cite{nonpertQED3} 
In particular, in the same way as it happens in QCD, the Lagrangian of QED$_3$ admits a non-trivial Chern-Simons (CS) term\cite{CST} which manifests itself as a gauge boson mass of topological nature, thus allowing for the possibility of time reversal and (generalized) parity breaking, fractional statistics and so on (see, for instance, Ref.~\refcite{revCS}).

In this work we propose that, for some planar systems effectively described by QED$_3$ in the presence of an external in-plane magnetic field, it occurs a mechanism which manifests itself as the generation of a non-dissipative electric current along the direction of the magnetic field. 
Such a current has a topological nature, and may be regarded as the 
analogue for a  bi-dimensional system of an effect proposed in the context of the quark gluon plasma produced in heavy ion collisions, known as chiral magnetic effect (CME).\cite{CME} 
The effect we describe is not related with the spin of charge carriers, but pseudospin. 
We refer to it as pseudochiral magnetic effect (PCME).

\section{The PCME in graphene}

The CME  is characterized by the interaction between the topological gauge fields and the fermions (in this case, deconfined quarks), that causes a flip in the chirality of the latter, generating domains of homogeneous chirality. 
Because in (3+1)-dimensions, chirality in the massless limit corresponds to helicity, a relation between the directions of spin and momentum is established, and in the presence of an external uniform magnetic field, the magnetic alignment of spins the spins with it 
generates an electric current in the field direction. 

In the massless Dirac theory, where $\gamma^5$ commutes with the Hamiltonian, the chirality quantum number is a conserved quantity and it is possible to define a representation where the spinors that describe the quasiparticle excitations are eigenstates of $\gamma^5$. 
In (2+1)-dimensions it is not possible to perform rotations around the direction of the momentum and the spin operator loses its usual physical meaning.
In other words, the concept of helicity related to Lorentz group and real space rotations is meaningless. 
However, one can still construct an operator that commutes with the Hamiltonian and has the chiral conserved quantum number as its eigenvalue. 
In analogy with the spin in (3+1)-dimensions, this is called pseudospin operator and corresponds to an internal symmetry rather than to spatial symmetry.\cite{Gusynin:2007ix}
In this case, the eigenstates of the $\gamma^5$ operator are the spinors that, in the case of monolayer graphene, represent the quasiparticle excitation at the two inequivalent Dirac points in the first Brillouin zone  of the graphene honeycomb, $K_+$ and $K_-$, and therefore the pseudohelicity can be seen as a flavor label corresponding to each one of the Dirac points. 
Considering the prescription described above, a breaking of the pseudohelicity symmetry in planar systems corresponds to an imbalance between the inequivalent Dirac points, which can be, for instance, by generating different effective masses for each one of them.

\begin{figure}[t!]
\includegraphics[width=\textwidth]{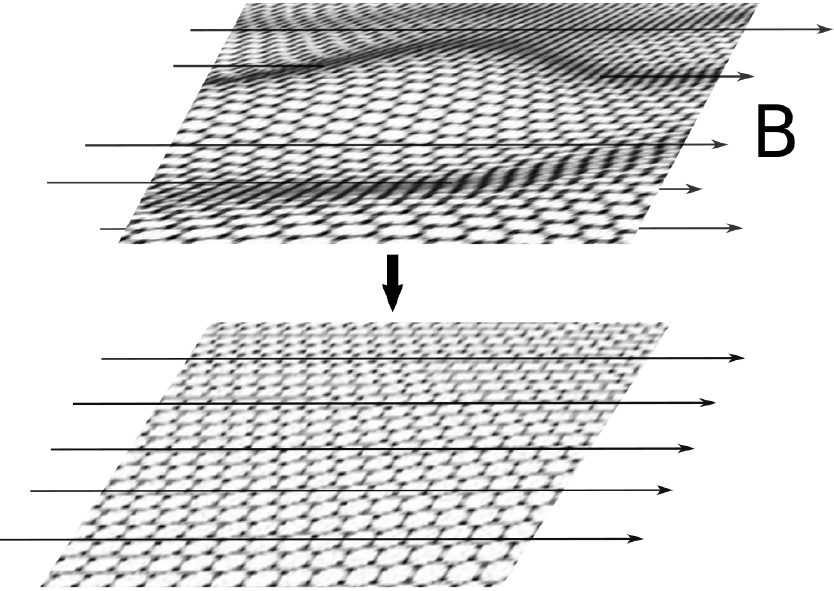}
\caption{A membrane of corrugated graphene in an in-plane magnetic field is idealized as a flake of pristine graphene where charge carriers from different sublattices have different effective masses.}
\label{fig1}
\end{figure}

Electrons in crystal lattices behave like quasiparticles, which means that their interactions can be represented by effective masses.  
Generally speaking, in graphene-like materials,  beyond the free electron picture, masses or gaps can be opened through a variety of external perturbations, e.g.\ strong enough magnetic fields, but also through mechanical distortion of the underlying lattice structure.\cite{Geim:2013}
A total of 36 gap-opening instabilities of the Dirac type in the spin, valley and superconducting channels have been considered in graphene and graphene-like structures (see Ref.~\refcite{chamon} and references therein). 
To establish the pseudohelicity symmetry breaking, we are particularly interested in mechanisms that generate different masses for the inequivalent Dirac points. 
A physical realization of this prescription is given in Ref.~\refcite{boron-nitride}, where the authors propose to place the graphene membrane over a hexagonal boron nitride layer, that is conformed in such a way that its lattice coincides with the graphene honeycomb lattice, generating a different effective mass for the charge carriers from different points $K_+$ and $K_-$. Such an inequivalence has proven to be valid, for instance, for strained graphene, where pseudomagnetic fields are generated in terms of effective masses~\cite{strain}. 
In fact, the effect of intrinsic curvature of the graphene membrane at the level of the equations of motion of charge carriers can be considered as if these particles develop an effective masses in flat space~\cite{curvature}.
In the same way, in this work we consider a generalized deformed graphene layer aligned with a magnetic field permeating it.
Schematically shown in Fig. \ref{fig1}, the deformations and distortions of the honeycomb lattice can be described as a perfectly planar system by the inclusion of Haldane masses.

Representing the pseudohelicity breaking in terms of a field theory, we look for a combination of masses in the Lagrangian, among all those allowed for such systems, that results in different masses for each eigenstate of the chiral operator. 
The particular choice of the masses proportional to $\gamma^3$ and $\gamma^3\gamma^5$ generates this pseudochirality imbalance, still keeping the Lagrangian invariant under the pseudochiral transformation $\psi \to e^{i \theta\gamma^5}\psi$.  
An interesting feature about the mass proportional to $\gamma^3\gamma^5$ is that it generates a CS term in the gauge sector of the Lagrangian, as well as a CS term generates this mass term in the fermion Lagrangian.\cite{CH-DW}
Therefore, the chosen mass term structure reinforces our analogy with the CME in QCD, where the CS term is also responsible for the breaking of helicity symmetry.

\section{Lagrangian with planar magnetic field}

Completing the analogy, we require a Lagrangian that includes a magnetic field with field lines pointing along the graphene plane. 
Unlike the case of quantum Hall effect with a magnetic field perpendicular to the graphene sheet, the gauge field that generates this in-plane magnetic field can depend on the coordinate perpendicular to the plane. 
Besides brane-world scenarios, where gauge invariance is guaranteed by construction (see Ref.~\refcite{Kotikov:2013eha} and references therein), an appropriate treatment is to extend the theory to a (3+1)-dimensional space--time with a compactified dimension, and then perform a strict dimensional reduction. 
An extended dimensionality  is a more realistic scenario in the sense that graphene indeed has an effective thickness besides the corrugations. 
However, if we want to consider the intersection of this external classical gauge field with the plane directly, there is no SO(1,3) gauge symmetry anymore, the $A_3^{\mathrm{ext}}$ component decouples from the other gauge vector components and now the gauge components $A_0^{\mathrm{ext}}$, $A_1^{\mathrm{ext}}$, $A_2^{\mathrm{ext}}$ are constrained to an  SO(1,2) symmetry \cite{Abreu:2001dq}. 

This scenario is similar to what is observed in high temperature dimensional reduction, where the $A_0$ component decouples from the other gauge vector fields.\cite{Andersen:1995ej}
The correct way to deal with this situation is to choose a gauge which does not depend on the perpendicular coordinate.
In this case the dimensional reduction is guaranteed and therefore we can treat the system as a (2+1)-dimension one scketched in Fig. \ref{fig1}.

Considering an external magnetic field  homogeneous in space and time,
the only gauge component that generates the planar magnetic field is $A_3^{\mathrm{ext}}$, since contribution from other vector components must depend on the dynamics of the transverse dimension. 
 If the external magnetic field points in the $x$-direction, we choose the Landau-like gauge $A^{\mathrm{ext}}_3=-By$, where the curl of this gauge field generates the in-plane magnetic field and its coordinate dependence is also in the plane.\footnote{Hereafter we use for coordinates the notation $(r^0,r^1,r^2)=(t,x,y)$.}
 The other external electromagnetic components are still gauge invariant.

The resulting general chiral invariant Lagrangian, in Minkowski space, is:
\begin{eqnarray}
{\cal L}_F &=& \bar\psi\big[i\slashed{D}
+(eA^{\mathrm{ext}}_3-m_3)\gamma^3  
-m_o\gamma^3\gamma^5\big]\psi\;,
\end{eqnarray}
where 
$D=(\partial_0-i\mu, v_F \bm{\nabla})$, $e$ is the fundamental charge, $v_F$ is the Fermi velocity and $\mu$ the chemical potential.
The fields $\psi$ are 8-component spinors which correspond to the direct product spin$\otimes$pseudospin. As we do not include spin interactions, spin label will not be treated explicitly in what follows, but must be taken into account.
For simplicity we will set $v_F=1$ or, in other words, $v_F$ will be the new speed of light.
Notice that because we describe the propagation of negative charge-carriers (quasiparticles and not holes), the chemical potential must be positive.

Recalling the discussion in the previous section, the presence of a CS term automatically generates the mass $m_o$ in the fermion Lagrangian. This mass is referred to in literature as the {\it odd mass}\cite{CH-DW} regarding its parity non-preserving character.
The mass term $m_3$ reproduces the asymmetry between the sub-lattices generated by, for example, placing the graphene  membrane on top of a hexagonal boron nitride layer.\cite{boron-nitride}\footnote{
In the Dirac representation, the mass $m_3$ corresponds to the standard Dirac mass.}

\section{Green Function}

In the Weyl representation for the gamma matrices, the charge carriers Lagrangian can be separated in two chiralities: 
\begin{equation}
{\cal L}_F=  \sum_{\chi=\pm} 
\bar\psi_\chi\left[i\slashed{\partial}+\mu\gamma^0
+(eA^{\mathrm{ext}}_3 - m_\chi)\gamma^3
\right]\psi_\chi\;,
\end{equation}
with the fields and masses defined as 
$\psi_\pm=\frac{1}{2}(1\pm\gamma^5)\psi$ and $m_\pm = m_3\pm m_o$.
The Green function in configuration space can be written in terms of the combination of the Green function of each chirality as
\begin{equation}
G(r,r') = \frac{1}{2}(1+\gamma^5) G_+(r,r') +\frac{1}{2}(1-\gamma^5) G_-(r,r'),
\label{G}
\end{equation}
where  the chiral Green functions are defined as
\begin{equation}
G_\pm(r,r')=\langle 
r|\frac{i}{\slashed{\Pi} +(eA_3^{\mathrm{ext}}-m_\pm)\gamma^3}|r'\rangle\,,
\end{equation}
with $\Pi=(i\partial_0+\mu ,i \bm{\nabla})$.
Since the operators $\Pi_0$ and $\Pi_1$ commute with the other 
operators involved, $\Pi_2$ and $A_3^{\mathrm{ext}}$, we can introduce a set of eigenstates  $|k^0\rangle$ and 
$|k^1\rangle$.
The chiral Green functions, then, can be written as 
\begin{eqnarray}
G_\pm(r,r') &=& \int \frac{d^2k_\parallel}{(2\pi)^3}~e^{-ik_\parallel\cdot 
(r-r')}
\langle y| \frac{
\slashed{K}_\parallel+\Pi_2\gamma^2 
+(eA_3^{\mathrm{ext}}-m_\pm)\gamma^3}{iH_\pm}
|y'\rangle,
\label{Gc-2}
\end{eqnarray} 
where we have introduced the Hamiltonian in the 
proper-time method, defined as 
\begin{equation}
H_\pm\equiv-{K_\parallel}^2+{\Pi_2}^2+(eA_3^{\mathrm{ext}}-m_\pm)^2 + ieB\gamma^2\gamma^3\;,
\end{equation}
and where the parallel momentum vectors are defined as $k_\parallel=(k^0,k^1,0)$ and 
$K_\parallel = (k^0+\mu,k^1,0)$. The last term of the above equation is the responsible for the effect we are proposing and is independent of the gauge choice if we deal with an extended (3+1)-dimensional theory with one compactified dimension.

As it is well-known, some care must be taken when dealing with a chemical 
potential using the proper 
time method in a uniform magnetic background.\cite{chodos}
The reason is that when the 
chemical potential is larger than the fermion mass, 
the propagator must be regularized in a certain $\mu $-dependent 
way. 
The time-ordered regulator is defined setting $k^0\to k^0(1+\varepsilon)$. 
As a consequence, in the denominator of Eq. (\ref{Gc-2}), ${K_\parallel}^2\to 
{K_\parallel}^2+i\varepsilon 2k^0K^0_\parallel$.
So, in order to express the denominator as an integral 
of an exponential term, the convergence will 
be determined by the sign of $k^0(k^0+\mu )$.
This can be written in a simple way as 
\begin{equation}
\frac{1}{iH_\pm +\varepsilon 2k^0K^0_\parallel}
=\int_{-\infty}^{\infty}ds ~r_s(k^0K^0_\parallel)~e^{-isH_\pm}\;,
\end{equation}
where the regulation function, defined as
\begin{equation}
r_s(\zeta)=\theta(s)\theta(\zeta)e^{-s\varepsilon}
-\theta(-s)\theta(-\zeta)e^{ s\varepsilon},
\label{rs}
\end{equation}
ensures the correct convergence of the integral in the proper time.
Now, the  Schwinger 
proper-time method \cite{schwinger} can be performed in the usual way by identifying the 
$y$-coordinate states as $\langle y|=\langle y(s)|$ and 
$|y'\rangle=|y'(0)\rangle$.
We write the resulting chiral  Green functions in an appropriate 
form:
\begin{eqnarray}
G_\pm(r,r') &=&   \int\!\!\!\frac{d^3 k}{(2\pi)^3}
e^{-ik\cdot(r-r')} \tilde G\left(\! k;\frac{1}{2}(y+y')eB+m_\pm
\!\right),
\nonumber\\
\label{Gc3}
\end{eqnarray}
where 
\begin{eqnarray}
 \tilde G(k;\xi &)&= \int_{-\infty}^{\infty} \!\!\!
ds~r_s(k^0K^0_\parallel)~
e^{is{K_\parallel}^2-i
\left[{k_2}^2+\xi^2\right]
\tan(eBs)/eB}
\nonumber\\&&
\left\{ \slashed{K}_\parallel\left[1 \!+\! \gamma^2\gamma^3 \tan(eBs)\right]
+\left[k_2\gamma^2 \!+\! \xi\gamma^3\right]\sec^2(eBs)\right\}\!\!.
\label{tildeG}
\end{eqnarray} 
The term $\xi = \frac{1}{2}eB(y+y')+m_\pm$ is a nonlocal factor along the direction perpendicular to the magnetic field on the plane.

\bigskip
The  effects of a thermal bath are introduced  by the replacements
$ k^0\to i\omega_n$, and $\int dk^0\to 2\pi iT\sum_n$, where $\omega_n=(2n+1)\pi T$ are the Matsubara frequencies, and the time 
component $t\to -i\tau$ is now compactified in the region 
$0\leq\tau\leq1/T$ .
The regulator for the proper-time can be introduced with the same analysis and the replacement is 
\begin{equation}
r_s\big(k^0(k^0+\mu)\big)\to r_s(\omega_n\mu),
\label{rT}
\end{equation}
In the case of zero temperature in Euclidean space, the procedure follows the same line: $k^0\to -ik_4$ introducing the regulator in $r_s(-k_4\mu)$.

\section{Current densities}

Let us now calculate the currents. A current density is defined as
\begin{equation}
(-e)\langle \bar\psi\Gamma\psi\rangle
= e\,\mathrm{tr}\Gamma G(r,r) =  j_\Gamma(y)   \;,
\label{currents}
\end{equation}
where the trace is taken over spin and pseudospin indexes and
where the operator $\Gamma$ can be $\gamma^\mu, \gamma^\mu\gamma^5,
\gamma^3, \gamma^3\gamma^5$. 
The dependence on $y$, explicitly indicated in Eq.~(\ref{currents}),
appears due to the nonlocal term in the chiral Green function in Eq.~(\ref{Gc3}) and (\ref{tildeG}).

Our model, even in the absence of a magnetic field, presents some non-vanishing expectation values: the particle number density $n=\langle\psi^\dag\psi\rangle$ due to the chemical potential; the condensates $\sigma_3 = \langle \bar\psi\gamma^3\psi\rangle$ and $\sigma_o = \langle \bar\psi\gamma^3\gamma^5\psi\rangle$ generated through the masses $m_3$ and $m_o$, respectively; and the chiral number density  $n_5=\langle\psi^\dag\gamma^5\psi\rangle$, due to the combination of chemical potential and masses.
The presence of the uniform external magnetic field along the $x$-direction 
catalyzes the generation of two other currents along the same direction: an electric current $j_x  = -e\langle\bar\psi\gamma^1\psi\rangle$ and an axial current $ j_{5x} = -e\langle\bar\psi\gamma^1\gamma^5\psi\rangle$, which is expected in analogy with the CME produced in QCD.
We explore these currents in more detail below.

Following the expressions for the Green function in Eq. (\ref{G}), (\ref{Gc3}) 
and (\ref{tildeG}), tracing over spin and pseudospin, we can write the electric induced current density and the axial current density as \begin{eqnarray}
j_{x}(y) &=& j(y-y_+) - j(y-y_-),
\\
 j_{5x}(y) &=& j(y-y_+) + j(y-y_-),
\end{eqnarray}
with $y_\pm = -m_\pm/eB$, and where the function $j$ reads
\begin{eqnarray}
j(\eta) &=&
2\frac{e^2B T}{2\pi}\sum_n  \int_{-\infty}^{\infty} 
ds~r_s(\omega_n\mu)~ (\omega_n-i\mu)
\left[\frac{\tan(eBs)}{eBs}\right]^{1/2}
\nonumber\\&&
\exp\left(-i\left[s(\omega_n-i\mu)^2+eB\tan(eBs)\eta^2\right]\right).
\end{eqnarray}
Note that the role of the spin in this case is simply to duplicate the degrees of freedom and it is not involved in any interaction.

It is possible to rotate the proper time integral in the complex plane at finite temperature.  
Following the restrictions imposed by the regulator $r_s$ defined in Eq. (\ref{rs}) and (\ref{rT}), we separate the proper time integral into the negative and  positive ranges of integration.  
If $\omega_n^2>\mu^2$, the integrals can be enclosing the contour in the lower complex plane.
If $\omega_n^2<\mu^2$, the contour of integration encloses in the upper  complex plane. 
Both results are different, so, if we want a single expression for all the Matsubara frequencies, we must restrict ourselves to the condition $\pi T > \mu$.
The function $j$ in this case can be written as
\begin{eqnarray}
j(\eta) &=&
i\frac{e^2B T}{\pi}\sum_n  \int_{0}^{\infty} 
ds~(\omega_n-i\mu)
\left[\frac{\tanh(eBs)}{eBs}\right]^{1/2}
\nonumber\\&&
\exp\left(-\left[s(\omega_n-i\mu)^2+eB\tanh(eBs)\eta^2\right]\right).
\label{j_rotated}
\end{eqnarray}

The integral above is highly suppressed by the exponential term $e^{-s(\omega_n^2-\mu^2)}$, where the range $s<(\pi T)^2-\mu^2$ dominates. 
If $|eB|<(\pi T)^2-\mu^2$, the product $eB s$ is too small in the relevant integration region.
Then, we can make  the approximation $\tanh(eBs)\approx eBs$ simplifying  enormously (\ref{j_rotated}), allowing us to integrate the proper time. 
After summing all the Matsubara frequencies we get this simple result: 
\begin{equation}
j(\eta) \approx -\frac{e^2B}{2\pi}\left[n_F(|eB\eta|-\mu)-n_F(|eB\eta|+\mu)\right],
\end{equation}
being $n_F$ the Fermi-Dirac distribution.

The total electric current along the $x$ direction,
following \refcite{Miransky:2015ava},  is then
\begin{equation}
I_x =\int_{-L_y/2}^{L_y/2} dy\left[j(y-y_+)-j(y-y_-)\right],
\end{equation}
being $L_y$ the size of the plane in the $y$ direction.

\section{Conclusions}

In conclusion, we present here a new transport mechanism for systems represented by (2+1)-dimensional quantum electrodynamics in the presence of an external in-plane uniform magnetic field.
The systems considered present topological deformations represented by a 
CS term which induces a mass term proportional to $\gamma^3\gamma^5$ as well as sub-lattice deformations represented by a mass gap proportional to $\gamma^3$.
We showed that  the presence of an external in-plane magnetic field in such configuration generates an electric current along the field lines. 
Such a phenomenon can be regarded as an analogue of the CME proposed for QCD. 

On one hand, the PCME has potential technological applications on material physics. 
On the other hand, because heavy ion collisions environment possess intrinsic ambiguities on the observables  related to the non-trivial vacuum of QCD, the condensed matter systems offer a more controlled environment that can provide valuable insight on the knowledge of the QCD vacua. 
Our expectations are that phenomena like the one we describe can shed light on the issue.

\section*{Acknowledgements}
The Huitzil collaboration acknowledges J.~A.~Helayel-Neto for valuable discussions, W. Bietenholz for carefully reading this manuscript.
We also acknowledge hospitality of UMSNH (Mexico) and UBB (Chile), where parts of this work were carried out,  and La Porfiriana for the inspiration. 
AJM acknowledges CONACyT-Mexico under grant number 128534 and SECITI/CLAF under grant 023/2014.
AR acknowledges support from CIC-UMSNH grant No. 4.22 and CONACyT-Mexico under grant number 128534. 
CV acknowledges support from FONDECYT under grant numbers  1150847, 1130056 and  1150471.

\end{document}